\documentstyle[aps,preprint,epsfig]{revtex}

\tighten
\begin{document}

\draft

\title{Transition Rates between Mixed Symmetry States: 
       First Measurement in $^{94}$Mo} 

\author{N. Pietralla$\,^1$, C. Fransen$\,^1$, D. Belic$\,^2$, 
        P. von Brentano$\,^1$, C. Frie\ss{}ner$\,^1$, 
        U. Kneissl$\,^2$, A. Linnemann$\,^1$, 
        A. Nord$\,^2$, H.H. Pitz$\,^2$, T. Otsuka$\,^3$, 
        I. Schneider$\,^1$, V. Werner$\,^1$, I. Wiedenh\"over$\,^{1}$\thanks{ 
Present address:Argonne National Laboratory, Argonne, Illinois 60439} }

\address{$^1\,$ Institut f\"ur Kernphysik, Universit\"at zu K\"oln, 
                50937 K\"oln, Germany}
\address{$^2\,$ Institut f\"ur Strahlenphysik, Universit\"at Stuttgart, 
                70569 Stuttgart, Germany}
\address{$^3\,$ Department of Physics, University of Tokyo, 
      Hongo, Bunkyo-ku, Tokyo 113-0033, Japan }
 
\date{\today}

\maketitle

\begin{abstract}
The nucleus $^{94}$Mo was 
investigated using a powerful combination of $\gamma$-singles photon 
scattering experiments and $\gamma\gamma$-coincidence studies 
following the $\beta$-decay of $^{94m}$Tc. 
The data survey short-lived $J^\pi = 1^+,\,2^+$ states and 
include branching ratios, $E2/M1$ mixing ratios, lifetimes and 
transition strengths. 
The proton-neutron mixed-symmetry (MS) $1^+$ scissors mode and the $2^+$ MS 
state are identified from $M1$ strengths. 
A $\gamma$ transition between MS states was 
observed and its rate was measured. 
Nine $M1$ and $E2$ strengths involving 
MS states agree with the O(6) limit of the 
Interacting Boson Model-2 using the proton boson $E2$ charge as the only 
free parameter. 
\end{abstract}


\narrowtext

Enhanced magnetic dipole ($M1$) $\gamma$ transitions between low-lying 
states of heavy nuclei are of great interest \cite{RichRev}. 
Investigations were influenced by Lo\,Iudice and Palumbo 
\cite{LoPa78} predicting the scissors mode. 
Later on, Iachello predicted \cite{Iach81} enhanced $M1$ transitions 
between low-lying states within the proton-neutron (pn) 
version \cite{ArOt77} of the Interacting Boson Model (IBM-2). 
According to the IBM-2 approach an enhanced $M1$ strength is a general 
feature of a pn degree of freedom. 
The pn symmetry of an IBM-2 wave function is quantified by 
the $F$-spin quantum number \cite{OtAr78}. 
$F$-spin is the isospin for the elementary proton and neutron bosons. 
The IBM-2 predicts enhanced $M1$ transitions between states with 
$F$-spin quantum numbers $F_{\rm max}$ and $F_{\rm max}-1$. 
The latter are not fully symmetric with respect to the pn 
degree of freedom and are called mixed-symmetry (MS) states.

The recently proposed $Q$-phonon scheme 
\cite{Siem94,OtKi94,PiMi98} is an approximate scheme in the IBM. 
In this scheme the wave functions 
of the lowest symmetric and MS \cite{KiBu96,PiBa98,OtsQQm} states are 
approximated by simple expressions involving the proton and neutron 
quadrupole operators: 
\begin{eqnarray} 
\label{eq:21alsQs} 
|2^+_1\rangle \propto & Q_s|0^+_1\rangle & F=F_{\rm max}\\ 
\label{eq:22alsQsQs} 
|2^+_2\rangle \propto & \left(Q_sQ_s\right)^{(2)}|0^+_1\rangle  \hspace{0.8cm}
      & F=F_{\rm max}\\ 
\label{eq:2msalsQm} 
|2^+_{\rm ms}\rangle \propto & Q_{\rm m}|0^+_1\rangle & F=F_{\rm max}-1\\ 
\label{eq:1scalsQsQm} 
|1^+_{\rm sc}\rangle \propto & 
      \left(Q_sQ_{\rm m}\right)^{(1)}|0^+_1\rangle & F=F_{\rm max}-1\ . 
\end{eqnarray} 
Here, $Q_s = Q_\pi + Q_\nu$ is the symmetric sum of the proton and 
neutron boson quadrupole operators and $Q_{\rm m} = Q_\pi/N_\pi - 
Q_\nu/N_\nu$ is the orthogonal linear combination with 
$\langle 0^+_1|Q_s\cdot Q_{\rm m}|0^+_1\rangle = 0$. 
$N_\pi$ ($N_\nu$) denotes the number of proton (neutron) bosons. 
The $Q$-phonon scheme generalizes the bosonic phonon concept 
in vibrators. 
In contrast to that, the $Q$ operators do not have to obey the 
boson commutation relation. 
Furthermore, the $Q$ operators are applied to the true ground state, 
which can be correlated. 
The lowest $2^+$ MS state is interpreted as the MS one-$Q$-phonon 
excitation, which is orthogonal to the symmetric one-$Q$-phonon 
excitation, the $2^+_1$ state. 
The $1^+_{\rm sc}$ state is a MS two-$Q$-phonon state. 
The two-$Q$-phonon structure can be tested by measuring 
$E2$ strengths of decay transitions from the MS 
$1^+$ and $2^+$ states. 
Enhanced $M1$ transitions between states with $F$-spin quantum numbers 
$F_{\rm max}$ and $F_{\rm max}-1$ are expected 
to have matrix elements of the order of 1 $\mu_N$. 
In $\gamma$-soft nuclei one expects \cite{VaIs86}, for instance, 
the following enhanced $M1$ transitions: 
$1^+_{\rm sc}\rightarrow 2^+_2$ and  $2^+_{\rm ms}\rightarrow 2^+_1$.

In the early 1980ies Richter and co-workers discovered the 
MS $J^\pi = 1^+_{\rm sc}$ state in electron scattering 
($e$,$e^\prime$) experiments \cite{BoRi84} in Darmstadt. 
This discovery was supported by photon scattering 
($\gamma$,$\gamma^\prime$) experiments 
\cite{Berg84} in Stuttgart. 
Subsequent ($e$,$e^\prime$) \cite{RichRev} and systematic 
($\gamma$,$\gamma^\prime$) experiments \cite{KnPi96} accumulated 
knowledge about the $1^+$ scissors mode. 
This enabled systematic studies of the $M1$ excitation strength 
\cite{Piet95,PvNC95} and the excitation energy \cite{Piet98,Ende99} 
of the $1^+$ scissors mode including information on 
weakly deformed nuclei. 
Knowledge about other MS states is sparser. 
In some weakly deformed nuclei $J^\pi = 2^+$ MS states were identified from 
lifetime measurements (e.g. \cite{PiBa98,GaLe96}). 
Further information about MS states was deduced 
from inelastic hadron scattering cross sections (e.g. \cite{DeLeo}), 
from $E2/M1$ mixing ratios $\delta$ (e.g. \cite{HaIr84,MoGa88}) and 
electron conversion coefficients measured in $\beta$-decay studies (e.g. 
\cite{GiNa96}).

In this Letter we report on the identification of the $2^+$ MS state 
\underline{and} the $1^+$ MS state in $^{94}$Mo. 
We identify the MS states from measured $M1$ strengths. 
We discuss the decays of the observed MS states 
including the first measurement of a transition rate 
between MS states and the first $E2$ strength of the 
$1^+_{\rm sc}\to 2^+_1$ transition. 
This new information on MS states was accessible due to the new and powerful 
combination of a ($\gamma$,$\gamma^\prime$) 
experiment on $^{94}$Mo and a $\gamma\gamma$-coincidence measurement 
of transitions following the $\beta$-decay of $^{94}$Tc to $^{94}$Mo. 
Thereby, we combine the capability of two experimental techniques: 
a) the singles spectroscopy by resonant photon scattering providing 
lifetime and spin information and 
b) the clean off-beam spectroscopy of $\gamma\gamma$-coincidences 
of transitions following $\beta$-decay enabling the measurement of 
small $\gamma$ branches and multipole mixing ratios. 
In favored cases $\beta$-decay 
strongly populates highly excited low-spin states among which we 
identify MS states. 
From this new combination of techniques we obtain a richness of information 
on absolute transition strengths 
from MS states, which gives a new quality to the investigation of MS 
states.

\begin{figure}

\centerline{\epsfig{file=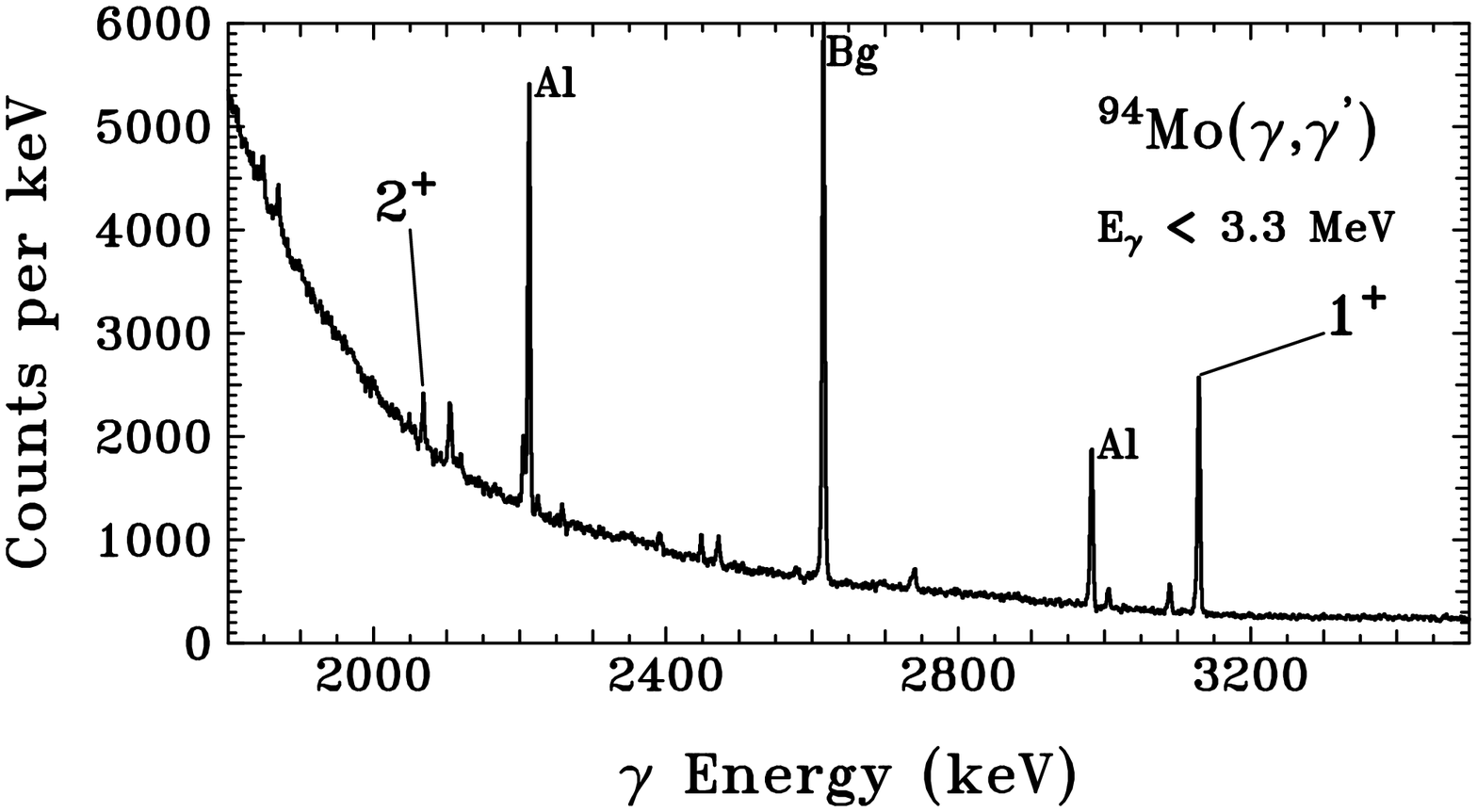,width=3.4in}}
\vspace*{10pt}
\caption{Photon scattering spectrum off \protect$^{94}$Mo in the 
         energy range of MS states. 
         At 2067 keV and 3129 keV we observe ground state transitions 
         of strongly excited \protect$2^+$ and \protect$1^+$ states.
         Photon scattering cross sections are measured relative to well 
         known \protect\cite{PietAl} cross sections in \protect$^{27}$Al, 
         which is irradiated simultaneously (marked "Al"). 
         "Bg" denote background lines.} 
\label{fig:NRFSpe}

\vskip1cm

\centerline{\epsfig{file=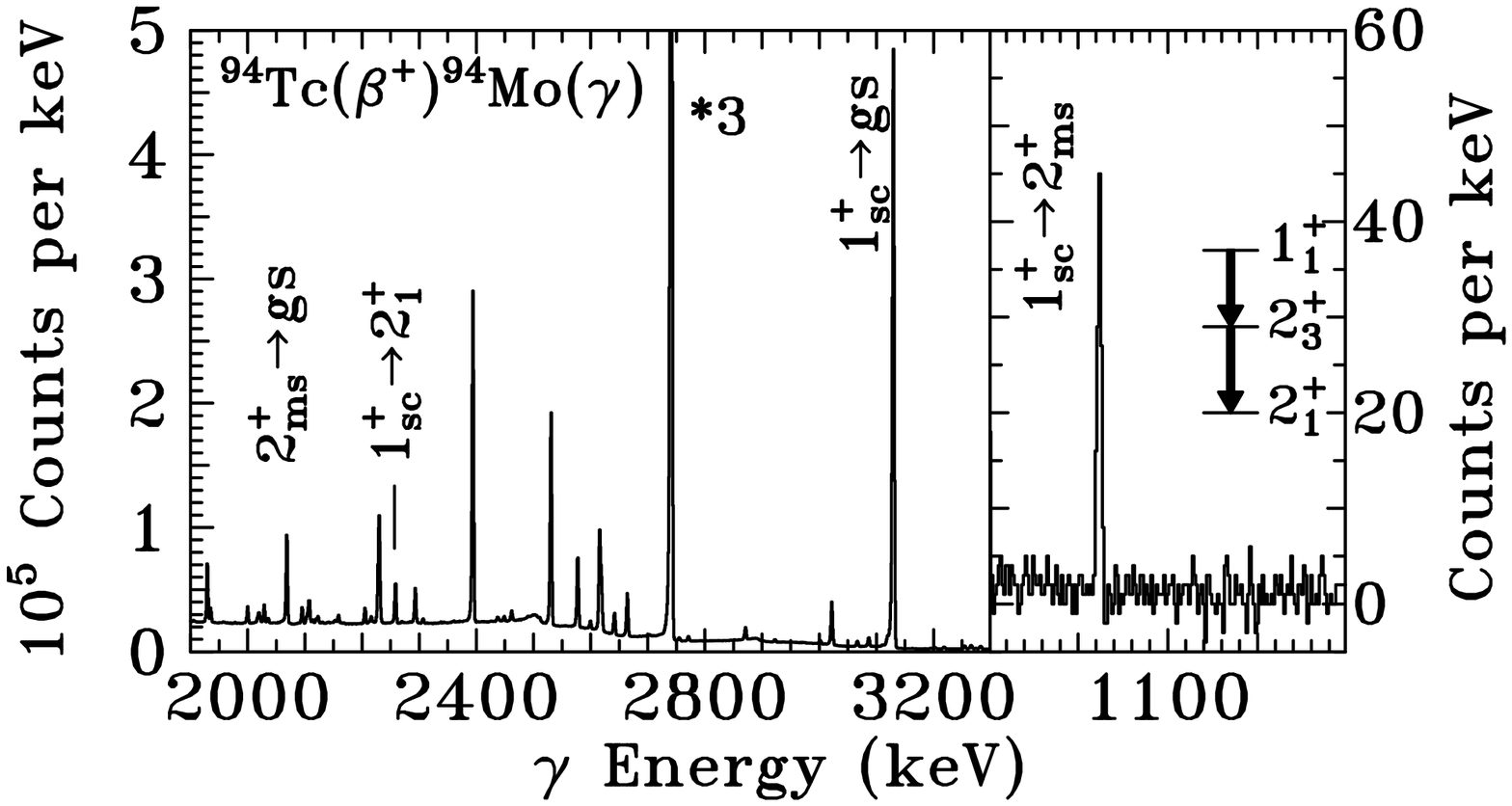,width=3.4in}}
\vspace*{10pt}
\caption{Left: Part of the observed spectrum of \protect$\gamma$-rays 
         following the 
         \protect$\beta$-decay of the \protect$J^\pi=(2)^+$ 
         low-spin isomer of \protect$^{94}$Tc populated in 
         the \protect$^{94}$Mo(\protect$p,n$) reaction. 
         High statistics and low background enable us to  
         observe weak decay branches 
         and to measure \protect$\gamma\gamma$-coincidences 
         for decays of MS states. 
         Right: Part of the \protect$\gamma\gamma$-coincidence spectrum 
         gated with the \protect$2^+_3 \to 2^+_1$ transition. 
         The coincident observation of the 1062 keV line establishes the 
         population of the \protect$2^+_3$ state at 2067 keV from the 
         \protect$1^+_1$ state at 3129 keV.} 
\label{fig:SpebetaSing}

\vskip1cm

\centerline{\epsfig{file=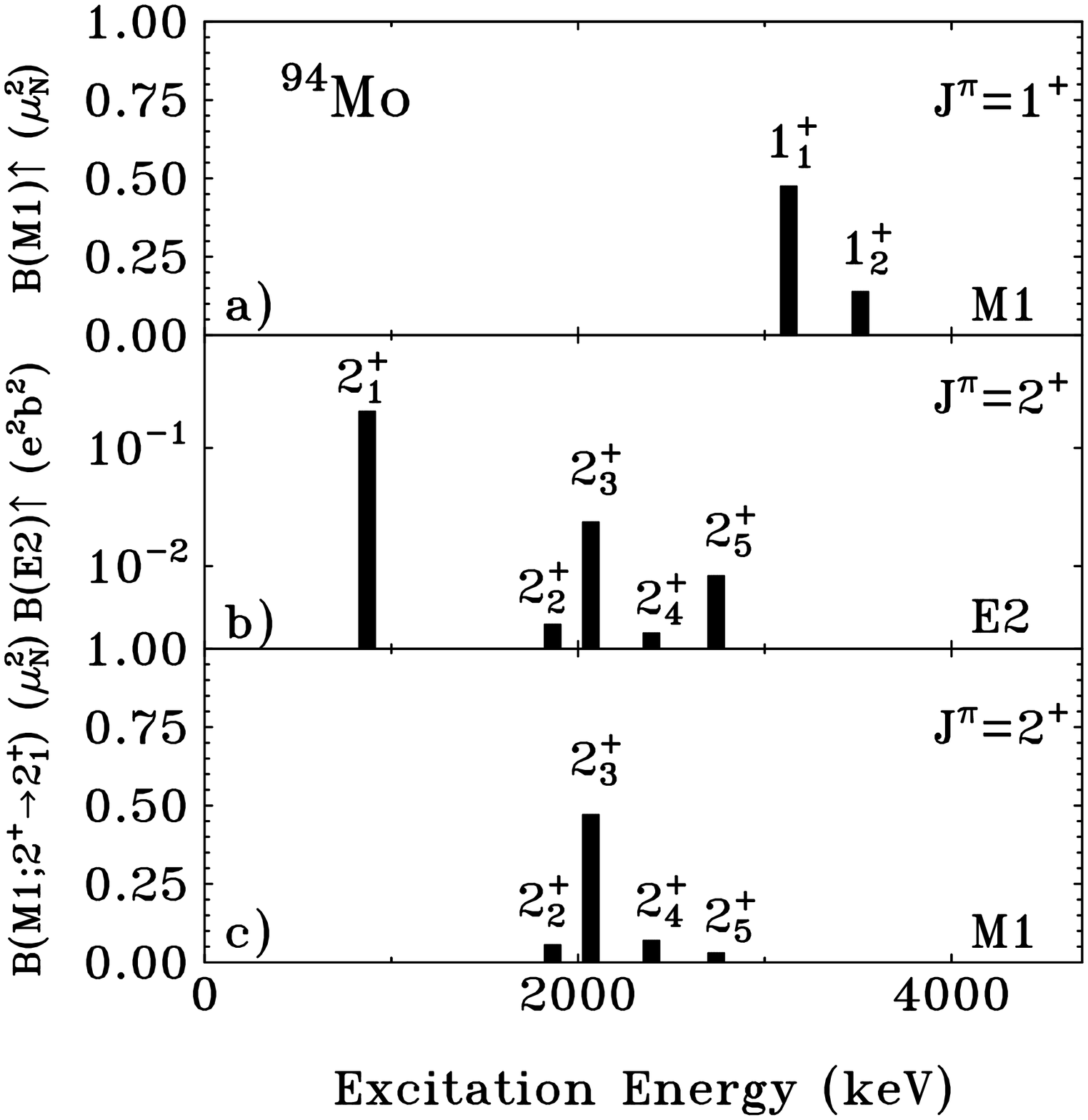,width=3.4in}}
\vspace*{10pt}
\caption{Measured \protect$M1$ and \protect$E2$ transition strengths 
         relevant for the identification of 
         \protect$1^+$ and \protect$2^+$ MS states in 
         \protect$^{94}$Mo. 
         Part a) and b) display the \protect$M1$ and the 
         \protect$E2$ excitation strength distributions 
         versus the excitation energies of the \protect$1^+$ 
         and \protect$2^+$ states. 
         Part c) shows the \protect$B(M1;2^+\rightarrow 2^+_1)$ 
         values for the four lowest non-yrast \protect$2^+$ states. 
         The \protect$1^+_1$ state is the main fragment of the 
         scissors mode. 
         The \protect$2^+_3$ state is the main fragment 
         of the \protect$2^+$ MS state.} 
\label{fig:ExcitDist}

\end{figure}

The photon scattering experiments were performed at the Dynamitron 
accelerator \cite{KnPi96} in Stuttgart. 
For bremsstrahlung production we used electron beams with 
energies of $E_e = 4.0$ MeV and $E_e = 3.3$ MeV. 
Figure \ref{fig:NRFSpe} shows the ($\gamma$,$\gamma^\prime$) 
spectrum off $^{94}$Mo taken at incident photon energies 
$E_\gamma < 3.3$ MeV. 
The $1^+$ state at 3129 keV with lifetime $\tau = 10(1)$ fs and 
the $2^+_3$ state at 2067 keV with $\tau = 60(9)$ fs are 
strongly excited. 
Below we interpret these states as the main 
fragments of the $1^+$ scissors mode and the $2^+$ MS state in $^{94}$Mo. 
We measured the photon scattering cross sections 
$I_{s,f} = g\pi^2\lambdabar^2\Gamma_0\Gamma_f/\Gamma$, where 
$g = (2J+1)/(2J_0+1)$ is a statistical factor and 
$\lambdabar = \hbar c/E_\gamma$ is the reduced wave length. 
$\Gamma$ and $\Gamma_{0}$ ($\Gamma_{f}$)  are the total level width and 
the partial decay width to the ground ($f$inal) state.

Decay intensity ratios $\Gamma_f/\Gamma$ were measured for 
low-spin states in $^{94}$Mo in a study of $\gamma$-rays 
following the $\beta$-decay of the $J^\pi = (2)^+$ low-spin 
isomer, $^{94m}$Tc. 
We produced $^{94m}$Tc nuclei in the center of the Cologne coincidence 
cube spectrometer using the reaction $^{94}$Mo($p$,$n$)$^{94}$Tc 
at an energy of $E_p = 13$ MeV. 
The beam was periodically switched on for 5 seconds to create 
activity and switched off for 5 seconds to observe singles 
$\gamma$ spectra and $\gamma\gamma$-coincidences of transitions 
following the $\beta$-decays. 
The singles spectrum between 1.9 MeV and 3.3 MeV 
is displayed in Fig. \ref{fig:SpebetaSing} on the left. 
The high counting rate, the low background in this off-beam 
measurement and the isotropy of the $\gamma$ radiation after 
$\beta$-decay enabled us to precisely determine the intensity 
ratios $\Gamma_f/\Gamma$. 
From our $\gamma\gamma$-coincidence data, we could place a 
1062 keV transition in the level scheme of $^{94}$Mo. 
The right part of Fig.\ref{fig:SpebetaSing} shows the 1062 keV 
transition in the background-subtracted $\gamma$ spectrum, which we 
observed in coincidence with the $2^+_3\to 2^+_1$ transition. 
The 1062 keV transition populates the $2^+_3$ state at 2067 keV directly 
from the $1^+_1$ state at 3129 keV. 
This transition is interpreted below as the 
$1^+_{\rm sc}\to 2^+_{\rm ms}$ transition between MS states. 
This is the first identification of such a transition.

Combining the measured photon scattering cross sections 
$I_{s,f}\propto \Gamma_0\Gamma_f/\Gamma$ and the decay intensity ratios 
$\Gamma_0/\Gamma$ from the $\beta$-decay experiment, 
we determined partial decay widths $\Gamma_f$, total level widths 
$\Gamma = \sum \Gamma_f$, lifetimes $\tau =\hbar/\Gamma = 1/\sum w_f$, 
and transition rates $w_f = \Gamma_f/\hbar$. 
The transition rates enable an unique identification of 
short-lived collective states. 
For the most intense $\gamma$ transitions we could determine 
$E2/M1$ mixing ratios $\delta^2 = \Gamma_{f,E2}/\Gamma_{f,M1}$ 
from the measured $\gamma\gamma$-angular correlations. 
Details will be given in a subsequent full length article. 
The measured, partial, single-multipolarity decay widths 
$\Gamma_{f,\pi\lambda}$ are proportional to the reduced 
transition strengths $B(\pi\lambda)$.

Figure \ref{fig:ExcitDist} shows measured $M1$ and $E2$ strengths which are 
relevant for the identification of $1^+$ and $2^+$ MS states. 
For the $2^+_{1,2}$ states the $E2$ excitation strengths have been taken 
from \cite{RaMa87,BaBa72}; all other data are from this work. 
The total $M1$ strength from the ground state to the 
$1^+$ states at 3129 keV and 3512 keV amounts to 
$\sum B(M1)\uparrow = 0.61(7)\ \mu_N^2$. 
The weighted average $1^+$ energy lies at 3.2 MeV. 
These data fit well into the systematics of the $1^+$ scissors mode 
observed so far: 
From the empirical formulae \cite{Piet95,Piet98,Ende99}, extracted from 
data on the $1^+$ scissors mode in the rare earth region, we expect the 
scissors mode in $^{94}$Mo at an excitation energy of 3.2 -- 3.5 MeV 
with a total excitation strength of $B(M1)\uparrow \approx 0.55 \ \mu_N^2$. 
The extrapolation of the empirical formulae agree 
with our observations. 
This is a strong argument that the $1^+_1$ state is the main fragment of the 
scissors mode ($1^+$ MS state) in $^{94}$Mo.

The $E2$ strength distribution, shown in Fig. \ref{fig:ExcitDist}b, is 
dominated by the $2^+_1$ state, which is the pn symmetric 
one-$Q$-phonon excitation. 
The $E2$ excitation strength of the $2^+_3$ state amounts to 
10\% of the $0^+_1\rightarrow 2^+_1$ strength. 
This is one order of magnitude more than the $E2$ excitation strength 
to the $2^+_2$ state, which is a symmetric two-$Q$-phonon state. 
The weakly-collective $0^+_1\to 2^+_3$ $E2$ transition 
suggests that the $2^+_3$ state is a one-$Q$-phonon excitation, 
in agreement with Eq. (\ref{eq:2msalsQm}). 
Part c) of Fig. \ref{fig:ExcitDist} shows the $M1$ transition 
strengths of the four lowest non-yrast $2^+$ states to the $2^+_1$ state. 
Only the $2^+_3$ state decays via an enhanced $M1$ transition to the 
$2^+_1$ state. 
The enhanced $2^+_3\to 2^+_1$ $M1$ transition and the weakly-collective 
$2^+_3 \to 0^+_1$ $E2$ transition agree with the MS interpretation 
for the $2^+_3$ state.

The $1_1^+$ state and the $2^+_3$ state can be 
described quantitatively as MS states in IBM-2. 
In order to reduce the number of free parameters, 
we compare the measured transition strengths to the 
predictions of the O(6) dynamical 
symmetry\footnote{We use the Ginocchio sum rule for \protect$B(M1)$ 
strength \protect\cite{Gino91} and the total strength 
\protect$\sum B(M1)\uparrow = 0.61(7)\,\mu_N^2$ which we 
observed below 4 MeV and we derive a fraction of 42(5)\% 
\protect$d$-bosons in the IBM-2 ground state wave function of 
\protect$^{94}$Mo. 
This large \protect$d$-boson content rules out the U(5) dynamical 
symmetry limit (no \protect$d$-boson in the ground state) for 
an adequate IBM-2 description of \protect$^{94}$Mo and favors the 
O(6) limit, which predicts a fraction of 33\% 
\protect$d$-bosons in the IBM-2 ground state of \protect$^{94}$Mo.}. 
These predictions are independent of any Hamiltonian parameters and 
are simple analytical expressions \cite{VaIs86}, which involve the 
boson numbers and the parameters of the transition operators, only. 
We consider the doubly-closed shell nucleus $^{100}$Sn as the core 
and, consequently, use $N_\pi = 4$ proton bosons and $N_\nu = 1$ 
neutron boson. 
We reduced the number of parameters in the transition operators further 
by restricting them to the proton parts alone: 
$T(M1) = \sqrt{3/4\pi} g_\pi\,L_\pi$ and  $T(E2) = e_\pi\,Q_\pi$. 
Here $L_\pi$ and  $Q_\pi$ are the standard proton angular momentum operator 
and the proton quadrupole operator in the O(6) limit ($\chi_\pi=0$). 
Moreover, we must assume the orbital value $g_\pi = 1\,\mu_N$ for the 
proton boson $g$-factor leaving the effective quadrupole boson charge 
$e_\pi = 9\ e$fm$^2$ the only adjustable parameter for the description 
of absolute $M1$ and $E2$ transition 
strengths\footnote{In a recent numerical IBM-2 calculation 
\protect\cite{PiBa98} for symmetric and mixed-symmetry states in 
the (\protect$ N_\nu=1$)-nucleus \protect$^{136}$Ba good agreement 
between theoretical and experimental \protect$E2$ transition strengths 
was obtained by using also a vanishing effective quadrupole neutron boson 
charge \protect$e_\nu =0$ and a comparably large effective 
quadrupole proton boson charge \protect$e_\pi =15.6\,e {\rm fm}^2$.}. 
Table \ref{tab:ExpvsIBM} summarizes the relevant spectroscopic information 
in comparison to the IBM-2 values in the O(6) dynamical symmetry limit. 
The data, including 9 transition strengths from the $1^+$ and $2^+$ 
MS states, are in reasonable agreement with the O(6) limit of the 
IBM-2 using the effective proton boson quadrupole charge $e_\pi$ 
as the only free parameter.

For $\gamma$-soft nuclei $M1$ transitions obey 
selection rules \cite{Piepid} with respect to the 
$d$-parity quantum number $\pi_d = (-1)^{n_Q}$, i.e., 
the number of $Q$-phonons $n_Q$ modulo two does not change. 
According to Eqs. (\ref{eq:21alsQs}-\ref{eq:1scalsQsQm}) 
the $M1$ transition from the $1^+_1$ state to the 
$2^+_1$ state is $d$-parity-forbidden \cite{Piepid} while the 
$1^+_1\rightarrow 2^+_2$ $M1$ transition is allowed. 
The measured ratio of the corresponding $M1$ strengths is 0.02, 
confirming the $d$-parity selection rule. 
A dominant $E2$ character of the $1^+_{\rm sc}\to 2^+_1$ transition in 
$\gamma$-soft nuclei was previously assumed for the interpretation of 
data for the nuclei $^{196}$Pt \cite{Bren96} and $^{134}$Ba \cite{MaPi96}. 
Our measurement supports the earlier assumptions.

Of particular interest is the comparison of the $E2$ strengths, which are 
interpreted in the $Q$-phonon scheme as the annihilation of the MS 
$Q$-phonon, $Q_{\rm m}$. 
According to Eqs. (\ref{eq:2msalsQm},\ref{eq:1scalsQsQm}) the MS 
$Q$-phonon, $Q_{\rm m}$, is annihilated in both the weakly collective $E2$ 
transitions 
$2^+_{\rm ms} \rightarrow 0^+_1$ and $1^+_{\rm sc} \rightarrow 2^+_1$, 
respectively. 
The ratio of the measured $B(E2)$ values is 
\begin{equation} 
\label{eq:QmRat}
\frac{B(E2;1^+_1 \rightarrow 2^+_1)}{B(E2;2^+_3 \rightarrow 0^+_1)}  
   = 0.7(3)\ .
\end{equation} 
Within the error this $B(E2)$ ratio is one. 
We conclude, that the $1^+_1$ state is a two-$Q$-phonon excitation of the 
ground state, built up by the coupling of the symmetric ($Q_s$) and 
the mixed-symmetric ($Q_{\rm m}$) $Q$-phonon operators. 
Analogously, one expects from Eqs. (\ref{eq:21alsQs},\ref{eq:1scalsQsQm}) 
collective $E2$ strengths for the 
$1^+_{\rm sc}\to 2^+_{\rm ms}$ and the $2^+_1\to 0^+_1$ transitions. 
In the present paper the transition rate of the $1^+_{\rm sc}\to 2^+_{\rm ms}$ 
transition was measured. 
This represents the first measurement of a transition rate between 
two MS states. 
Due to the too weak intensity of the $1^+_1\to 2^+_3$ transition  
the $E2/M1$ mixing ratio could not be measured. 
From the $d$-parity selection rules we expect a dominant $E2$ character 
of the $1^+_{\rm sc}\to 2^+_{\rm ms}$ transition. 
Assuming a vanishing $M1$ contribution, the ratio of the 
energy-reduced transition rates 
\begin{equation} 
\label{eq:Ratiowred} 
\frac{w_{1^+_1\to 2^+_3}/
   E_\gamma(1^+_1\to 2^+_3)^5}
      {w_{2^+_1\to 0^+_1}/
   E_\gamma(2^+_1\to 0^+_1)^5} 
   = 1.5(2) \ . 
\end{equation}
equals the corresponding $B(E2)$ ratio. 
We find indeed a collective $E2$ strength, which is comparable to the 
collective $2^+_1\to 0^+_1$ decay strength. 
This fact gives further support for the two-$Q$-phonon interpretation of the 
$1^+_1$ state in $^{94}$Mo. 
The weakly collective $E2$ transition from the $1^+_1$ state 
to the $2^+_1$ state and the probable collective $E2$ transition 
to the $2^+_{\rm ms}$ state represent -- besides the large $M1$ transition 
strengths -- new and independent observables for the collectivity of the 
$1^+$ scissors mode. 
It is interesting, that these observables are deduced from E2 properties, 
which are considered to be well described by the IBM.

We thank A. Fitzler, P. Matschinsky and H. Tiesler 
for help with the experiments. 
We acknowledge discussions with A. Gelberg, A. Giannatiempo, J.N. Ginocchio, 
F. Iachello, R.V. Jolos, A. Leviatan, A. Richter, and A. Zilges. 
This work was supported by the {\em DFG} under 
Contracts No. Br 799/8-2/9-1 and No. Kn 154/30 and 
by the {\em JSPS} under Grant-in-Aid 0604 4249.

\begin{table}[htb]

\caption{Comparison of measured transition strengths to the prediction of 
         the O(6) limit of the IBM-2, 
         where the $1_1^+$, $2^+_3$ states have MS. 
         The IBM-2 reproduces the dominant \protect$E2$ 
         character of the \protect$1^+\to 2^+_1$ 
         transition. 
         Many transition strengths and the transition rate \protect$w$ 
         between the MS states are reproduced on an absolute scale 
         using one free 
         parameter \protect$e_\pi = 9\,e{\rm fm}^2$ only.} 
\smallskip
\begin{tabular}{rdd}
Observable   & Expt. & IBM-2 \\ 
$B(M1;1_1^+\rightarrow 0^+_1) \ (\mu_N^2)$  & 0.16(1) & 0.16 \\
$B(M1;1_1^+\rightarrow 2^+_1) \ (\mu_N^2)$  & 0.007$^{+6}_{-2}$ & 0 \\
$B(M1;1_1^+\rightarrow 2^+_2) \ (\mu_N^2)$  & 0.43(5) & 0.36 \\
$B(M1;1_1^+\rightarrow 2^+_3) \ (\mu_N^2)$  & $<$0.05 & 0 \\
$B(M1;2^+_2\rightarrow 2^+_1) \ (\mu_N^2)$ & 0.06(2) & 0 \\
$B(M1;2^+_3\rightarrow 2^+_1) \ (\mu_N^2)$ & 0.48(6) & 0.30 \\
$B(M1;2^+_4\rightarrow 2^+_1) \ (\mu_N^2)$ & 0.07(2) & 0 \\
$B(M1;2^+_5\rightarrow 2^+_1) \ (\mu_N^2)$ & 0.03(1) & 0 \\
$w(1^+_1\rightarrow 2^+_3) \ ({\rm ps}^{-1})$ & 1.02(12) & 0.92 \\
$\frac{I_\gamma(E2)}{I_\gamma}(1^+\rightarrow 2^+_1) \ (\%)$ & 
  60$^{+12}_{-21}$ & 100 \\
$B(E2;0^+_1\rightarrow 2^+_1) \ (e^2{\rm fm}^4)$ & 2030(40)\tablenote{from 
references \protect\cite{RaMa87,BaBa72}} & 2333 \\
$B(E2;0^+_1\rightarrow 2^+_2) \ (e^2{\rm fm}^4)$ & 32(7)$^{\rm a}$ & 0 \\
$B(E2;0^+_1 \rightarrow 2^+_3) \ (e^2{\rm fm}^4)$ & 230(30) & 151 \\
$B(E2;0^+_1 \rightarrow 2^+_4) \ (e^2{\rm fm}^4)$ & 27(8) & 0 \\
$B(E2;0^+_1 \rightarrow 2^+_5) \ (e^2{\rm fm}^4)$ & 83(10) & 0 \\
$B(E2;2^+_2 \rightarrow 2^+_1) \ (e^2{\rm fm}^4)$ & 720(260) & 592 \\
$B(E2;4^+_1 \rightarrow 2^+_1) \ (e^2{\rm fm}^4)$ & 670(100)$^{\rm a}$& 592 \\
$B(E2;2^+_3 \rightarrow 2^+_1) \ (e^2{\rm fm}^4)$ & $<$150 & 0 \\
$B(E2;1^+_1 \rightarrow 2^+_1) \ (e^2{\rm fm}^4)$ & 30(10) & 49 \\
$B(E2;1^+_1 \rightarrow 2^+_3) \ (e^2{\rm fm}^4)$ & $<$690\tablenote{assuming 
pure \protect$E2$ character, the value is 
\protect$620(70)\,e^2{\rm fm}^4$.} & 556 \\
\end{tabular}
\label{tab:ExpvsIBM}

\end{table}

\begin{references} 
 

\bibitem{RichRev} A. Richter, Prog. Part. Nucl. Phys. {\bf 34}, 
                  261 (1995). 
 
\bibitem{LoPa78} N.LoIudice, F.Palumbo, Phys.Rev.Lett.{\bf 41}, 
                 1532 (1978). 

\bibitem{Iach81} F. Iachello, Nucl. Phys. {\bf A358}, 89c (1981); 
                 Phys. Rev. Lett. {\bf 53}, 1427 (1984). 

\bibitem{ArOt77} A. Arima {\em et al.}, 
                 Phys. Lett. {\bf B66}, 205 (1977). 
 
\bibitem{OtAr78} T.Otsuka, A.Arima, F.Iachello, 
                 Nucl. Phys. {\bf A309}, 1 (1978). 
 
\bibitem{Siem94} G.~Siems {\em et al.}, 
                 Phys.~Lett.~{\bf B320}, 1 (1994).

\bibitem{OtKi94} T.~Otsuka, K.H.~Kim, Phys.~Rev.~C {\bf 50}, R1768 (1994).

\bibitem{PiMi98} N. Pietralla {\em et al.}, Phys. Rev. C {\bf 57}, 150 (1998). 


\bibitem{KiBu96} K.-H. Kim {\em et al.}, in 
                 {\em Capture Gamma Ray Spectroscopy and Related Topics}, 
                 Budapest 1996, 
                 edited G. Moln\'{a}r {\em et al.}, 
                 (Springer, Budapest, 1998).  
 
\bibitem{PiBa98} N. Pietralla {\em et al.}, 
                 Phys.~Rev. C {\bf 58}, 796 (1998). 

\bibitem{OtsQQm} T. Otsuka {\em et al.}, in preparation. 

\bibitem{VaIs86} P. Van\,Isacker {\em et al.}, 
                 Ann. Phys. (NY) {\bf 171}, 253 (1986). 

\bibitem{BoRi84} D. Bohle {\em et al.}, 
                 Phys. Lett. {\bf B137}, 27 (1984).  

\bibitem{Berg84} U.E.P. Berg {\em et al.}, 
                 Phys. Lett. {\bf B149}, 59 (1984). 

 
\bibitem{KnPi96} U.~Kneissl, H.H.~Pitz, A.~Zilges, 
                 Prog. Part. Nucl. Phys. {\bf 37}, 349 (1996). 
 
\bibitem{Piet95} N.~Pietralla {\em et al.}, 
                 Phys.~Rev.~C {\bf 52}, R2317 (1995). 
 
\bibitem{PvNC95} P.von Neumann-Cosel {\em et al.}, 
                 Phys. Rev. Lett. {\bf 75}, 4178 (1995). 
 
\bibitem{Piet98} N.~Pietralla {\em et al.}, 
                  Phys.~Rev.~C. {\bf 58}, 184 (1998). 

\bibitem{Ende99} J. Enders {\em et al.}, 
                 Phys. Rev. C {\bf 59}, R1851 (1999). 

\bibitem{GaLe96} P.E. Garrett {\em et al.}, 
                 Phys. Rev. C {\bf 54}, 2259 (1996). 

\bibitem{DeLeo}  
                 R. De Leo {\em et al.}, Phys. Rev. C {\bf 53}, 2718 (1996). 

\bibitem{HaIr84} W.D. Hamilton, 
                 A. Irb\"ack, J.P. Elliott, 
                 Phys. Rev. Lett. {\bf 53}, 2469 (1984). 

\bibitem{MoGa88} G. Moln\'{a}r, R.A. Gatenby, S.W. Yates, 
                 Phys. Rev. C {\bf 37}, 898 (1988). 
 
\bibitem{GiNa96} A. Giannatiempo {\em et al.}, 
                 Phys. Rev. C {\bf 53}, 2770 (1996). 

\bibitem{PietAl} N. Pietralla {\em et al.}, 
                 Phys. Rev. C {\bf 51}, 1021 (1995). 

\bibitem{RaMa87} S. Raman {\em et al.}, 
                 At. Data Nucl. Data Tab. {\bf 36}, 1 (1987). 

\bibitem{BaBa72} J. Barrette {\em et al.}, 
                 Phys. Rev. C {\bf 6}, 1339 (1972). 

\bibitem{Gino91} J.N. Ginocchio, Phys. Lett. {\bf B265}, 6 (1991). 

\bibitem{Piepid} N. Pietralla {\em et al.}, 
                    Phys. Rev. C {\bf 58}, 191 (1998). 

\bibitem{Bren96} P. von Brentano {\em et al.}, 
                 Phys. Rev. Lett. {\bf 76}, 2029 (1996). 

\bibitem{MaPi96} H. Maser {\em et al.}, Phys. Rev. C {\bf 54}, R2129 (1996). 



\end{references}
\end{document}